\documentclass[conference]{IEEEtran}
\IEEEoverridecommandlockouts
\usepackage[letterpaper, margin=1in, top=0.75in, bottom=1.05in]{geometry}
\usepackage{multicol}
\usepackage{cite}
\usepackage{amsmath,amssymb,amsfonts}
\usepackage{algorithmic}
\usepackage{graphicx}
\usepackage{float}
\usepackage{subcaption}
\usepackage{caption}
\usepackage{textcomp}
\usepackage{xcolor}
\usepackage{booktabs}
\usepackage{titlesec}
\usepackage{orcidlink}
\usepackage{tikz}
\def\BibTeX{{\rm B\kern-.05em{\sc i\kern-.025em b}\kern-.08em
    T\kern-.1667em\lower.7ex\hbox{E}\kern-.125emX}}
\begin{document}

\title{TS-EoH: An Edge Server Task Scheduling Algorithm Based on Evolution of Heuristic\\

\thanks{
Yuqi Zhao* is the corresponding author. This work is supported by the National Natural Science Foundation of China (Nos. 62032016 and 61972292), the Postdoctoral Fellowship Program of China Postdoctoral Science Foundation (Nos. GZC20240571), and the China Postdoctoral Science Foundation (Nos. 2024M751052)}
}

\author{\IEEEauthorblockN{1\textsuperscript{st} Yatong Wang}
\IEEEauthorblockA{\textit{School of Computer Science } \\
\textit{Central China Normal University}\\
Wuhan, China \\
yatong\_wang@mails.ccnu.edu.cn}

\and
\IEEEauthorblockN{2\textsuperscript{nd} Yuchen Pei}
\IEEEauthorblockA{\textit{School of Computer Science} \\
\textit{Central China Normal University}\\
Wuhan, China \\
ycpei@ccnu.edu.cn}

\and
\IEEEauthorblockN{3\textsuperscript{rd} Yuqi Zhao*}
\IEEEauthorblockA{\textit{School of Computer Science} \\
\textit{Central China Normal University}\\
Wuhan, China \\
yuqizhao@ccnu.edu.cn}
}

\maketitle

\begin{abstract}
With the widespread adoption of 5G and Internet of Things (IoT) technologies, the low latency provided by edge computing has great importance for real-time processing. However, managing numerous simultaneous service requests poses a significant challenge to maintaining low latency. Current edge server task scheduling methods often fail to balance multiple optimization goals effectively. This paper introduces a novel task-scheduling approach based on Evolutionary Computing (EC) theory and heuristic algorithms. We model service requests as task sequences and evaluate various scheduling schemes during each evolutionary process using Large Language Models (LLMs) services. Experimental results show that our task-scheduling algorithm outperforms existing heuristic and traditional reinforcement learning methods. Additionally, we investigate the effects of different heuristic strategies and compare the evolutionary outcomes across various LLM services.
\end{abstract}

\begin{IEEEkeywords}
Task Scheduling, Edge Computing, Evolutionary Computing, Heuristic Algorithms
\end{IEEEkeywords}

\section{Introduction}
\label{1}
\noindent
With 5G and IoT adoption, edge computing's low latency for real-time processing is increasingly crucial. Edge servers can reduce the physical distance to end users, providing low latency, mobility, and location sensitivity. They offload real-time processing services from cloud servers, easing their task-handling burden\cite{b1}. Edge servers handle computation-intensive tasks for mobile devices, compensating for their limited computing power\cite{b2}. However, the rise of terminal devices and the growth of mobile internet challenges task scheduling and resource management for edge servers. Critical issues include efficient task scheduling, resource optimization, and service quality assurance with limited server resources.

Current edge server task scheduling research mainly includes heuristic algorithms (e.g., Ant Colony Optimization\cite{b11}) and reinforcement learning-based algorithms (e.g., Reinforcement Learning with Pointer Networks \cite{b4}). Heuristic algorithms have a self-organizing solid, self-learning, and self-adaptive capabilities. They can achieve globally optimal solutions and are robust, making them suitable for complex problems. However, parameter settings often influence their performance, and designing these heuristics requires significant manual effort and expertise\cite{b3}. Reinforcement learning-based algorithms require extensive training sets and substantial computational resources, which is unfeasible in practical scenarios.

\begin{figure}[t]
  \centering
  \includegraphics[width=0.40\textwidth]{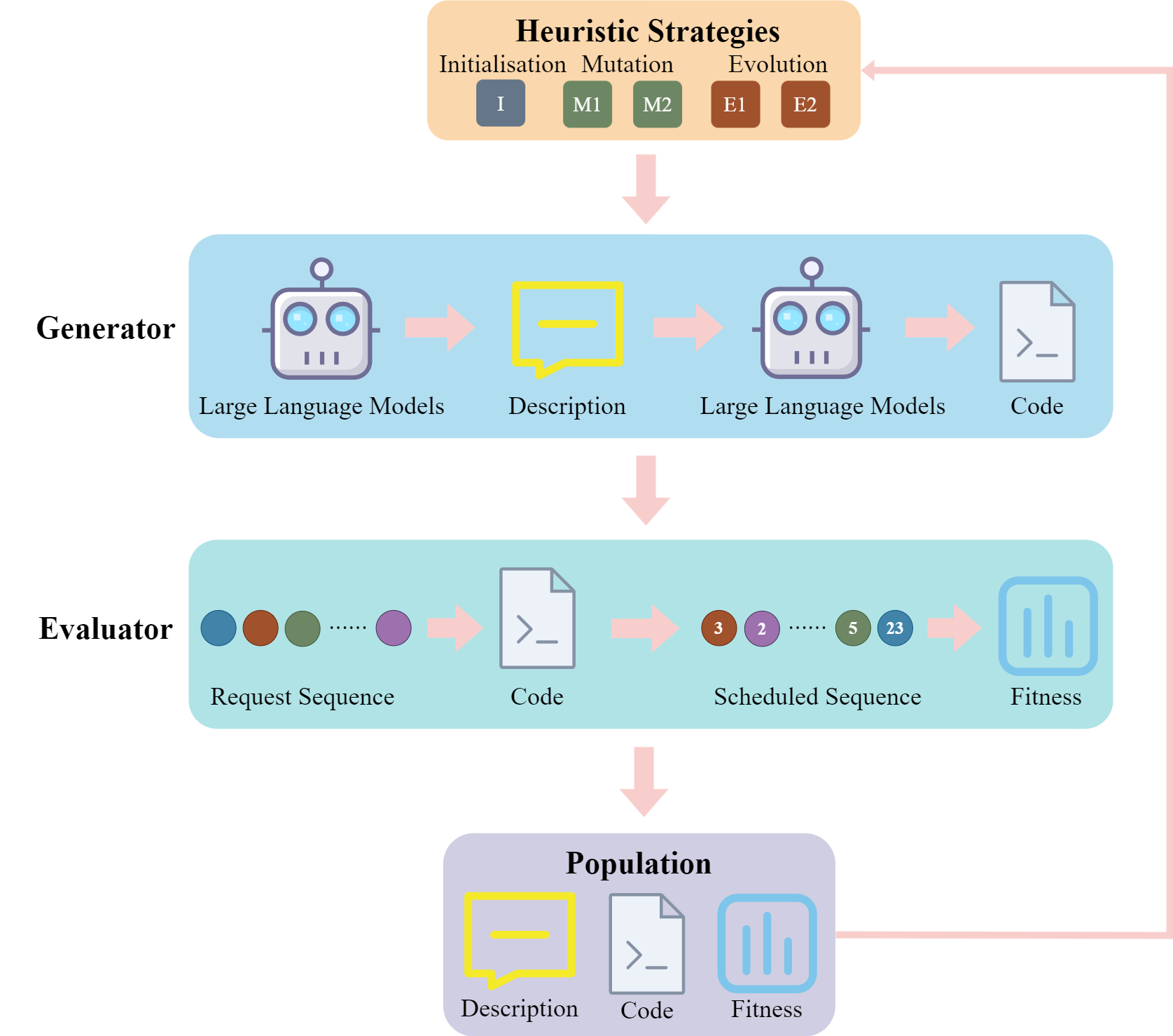} 
  \caption{The overall framework of edge server task scheduling algorithms based on EoH.}
  \label{fig1}
\end{figure}

The Evolution of the Heuristic (EoH) framework introduces a new approach to implementing heuristic algorithm design automatically. Using pre-trained data from LLMs, EoH guides LLMs through self-heuristic processes to evolve optimal scoring methods. This framework shows significant advantages in solving combinatorial optimization problems.

This paper proposes a \underline{t}ask \underline{s}cheduling scheme based on the \underline{EoH} framework (TS-EoH), combining EC theory and heuristic algorithms. The overall framework is shown in Fig. \ref{fig1}. We model service requests as task sequences and transform the task scheduling problem into a combinatorial optimization problem solvable by the EoH framework. LLMs-generated scoring methods evaluate each task sequence. These scoring methods focus on maximizing resource utilization and minimizing task running time.

The main contributions of this paper are:
\begin{itemize}
\item Introduction of a micro-parameter heuristic algorithm that reduces the impact of parameter settings in heuristic algorithms, avoiding complex parameter selection and achieving multi-objective optimization.
\item Refinement of EoH framework's automatic heuristics design capability to task scheduling problem, reducing dependence on expert knowledge and evolving optimal scoring schemes.
\item Comprehensive experiments are conducted to evaluate the effectiveness of the proposed algorithm using four LLMs services on three real datasets.
\end{itemize}

The remainder of this paper is organized as follows. Section \ref{2} introduces related work. Section \ref{3} details the implementation of the task scheduling scheme based on EC and heuristic algorithms. Section \ref{4} discusses experimental results on four LLMs services on three real datasets, including an ablation study. Section \ref{5} concludes this study.

\section{Related Work}
\label{2}
\noindent
To better understand the importance and innovation of our research, we first compare several task scheduling schemes based on heuristic algorithms in an edge computing environment. Following this, we introduce the background works of this paper.

\subsection{Heuristic task scheduling algorithm}
\label{2.2}
\noindent
Heuristic algorithms have shown significant advantages in solving large-scale combinatorial optimization problems\cite{b10}. Hence, they have a wide application scope in task scheduling problems of edge servers. Heuristic algorithms are mainly inspired by four kinds of heuristics that are Ant Colony Optimization (ACO), Particle Swarm Optimization (PSO), Genetic Algorithms (GA), and Hybrid Heuristic Algorithms (HHA).

ACO uses a pheromone mechanism to find the optimal path, making it suitable for handling scheduling issues with complex task dependencies and resource constraints. Also, it is robust and has global search capabilities. The AVE framework applies ACO algorithms to task scheduling to enhance efficiency and address the needs of assistance provided by other vehicles\cite{b11}. Wang and colleagues proposed the Load Balancing ACO (LBA) algorithm to solve task scheduling problems in a Mobile Edge Computing (MEC) environment with limited resources\cite{b12}. However, the ACO-based heuristic algorithms require extensive experimentation for parameter selection.

PSO is commonly used in dynamic scheduling environments, conducts a global search through the sharing of information among individuals within a swarm, and is capable of rapidly responding to arrived tasks and changes of resource status. \cite{b13} Presented a task scheduling scheme based on the PSO algorithm to enhance the resource utilization of cloud servers. \cite{b14} Introduced load balancing mutation Particle Swarm Optimization (LBMPSO), which considers multiple optimization objectives, including transmission costs and load balancing. However, it is not suitable for solving large-scale problems.

The emergence of GA \cite{b15} and HHA has provided new approaches for addressing multi-objective and multi-constraint optimization problems. GA seeks optimal solutions by simulating natural selection and genetic mechanisms. GA can be used for multi-objective optimization, such as minimizing task completion time and maximizing resource utilization. Hu et al. modeled the request scheduling problem as a dual decision problem based on mixed-integer nonlinear programming, using the Non-dominated Sorting Genetic Algorithm II (NSGA-II) for multi-objective optimization\cite{b16}. Yuan et al. proposed an improved NSGA-II for the intelligent workshop resource scheduling problem, addressing traditional GA's issue of slower search speed at the later stages by establishing a rank-based smoothing function and introducing a congestion mechanism\cite{b17}. HHA combines multiple heuristic algorithms to leverage their respective advantages. Wang et al. proposed a hybrid heuristic task scheduling algorithm that combines PSO and ACO \cite{b18}. Although these methods perfectly solve the problems of parameter settings and problem-scale limitation, they still require considerable effort in manual heuristic design.

\begin{figure*}[t]
  \centering
  \includegraphics[width=0.95\textwidth]{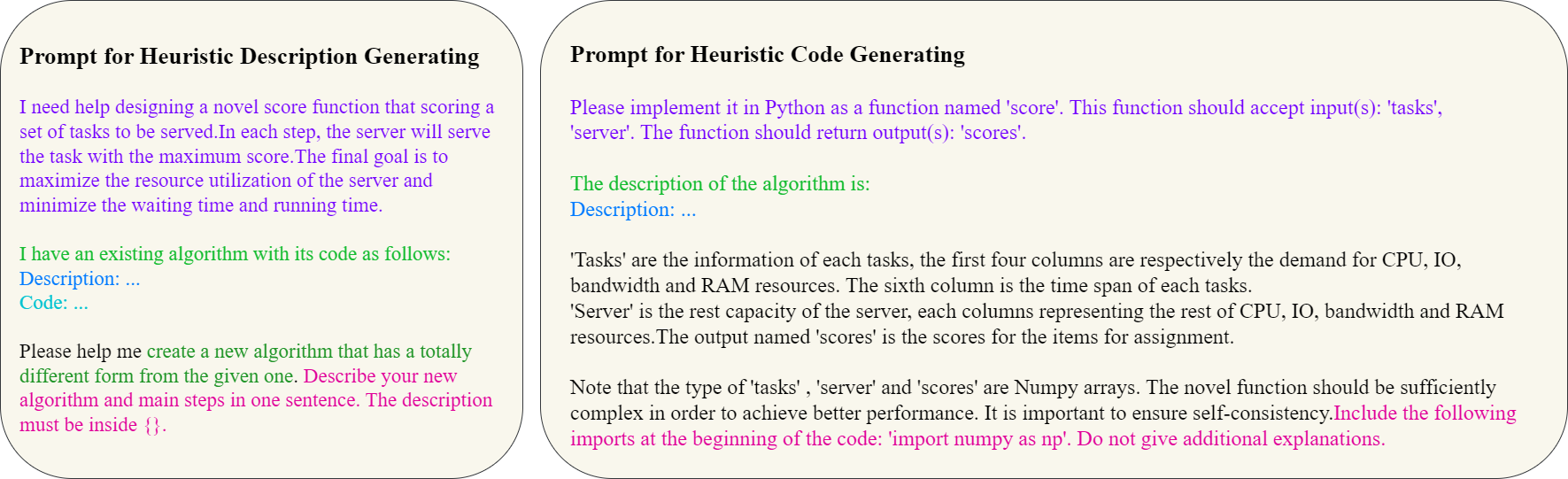} 
  \caption{Two examples of prompt engineering.}
  \label{fig6}
\end{figure*}

\subsection{Background works}
\label{2.1}
\noindent
\subsubsection{Serialization modeling scheme}
\label{2.1.1}
\noindent
In the service request scheduling algorithm RLPNet proposed by Zhao et al.\cite{b4}, the sequential characteristics of service request scheduling problems were noticed. Thus, they modeled it as a sequence-to-sequence (Seq2Seq) multi-objective optimization problem. The traditional Seq2Seq models exhibit poor scalability when dealing with the issue where the output sequence's length depends on the input sequence's length. Therefore, RLPNet employed a pointer network\cite{b5} for sequential modeling, which can handle variable-length input by outputting a probability distribution of the input sequence. This approach demonstrates significant advantages in reinforcement-learning-based task scheduling algorithms and provides reference value to many others.

\subsubsection{Automatic heuristic algorithm design based on LLMs}
\label{2.1.2}
\noindent
LLMs have been effectively employed in numerous practical scenes, showcasing their advanced capabilities in understanding and leveraging algorithmic structures and semantics. Wu et al. leveraged the capabilities of LLMs in capturing the structure and semantics of algorithms to propose, for the first time, an algorithm selection model based on LLMs\cite{b6}. This model was used to select the most suitable algorithm for solving a specific problem before execution. Shah et al. utilized the inference abilities of LLMs to generate possible heuristic search schemes to guide robots to make explorations purposefully\cite{b7}. However, all heuristic algorithm designs above based on LLMs require certain manual interventions.

Combining Evolutionary Algorithms (EAs) and heuristic algorithm designs based on LLMs greatly reduces the need for manual labor. The Evolution through Large Models (ELM) framework combines LLMs with EAs for the first time\cite{b8}, utilizing LLMs to generate mutation operators and embedding them into the process of algorithm evolution. EvoPrompt is an automatic prompt tuning architecture that combines the language processing capabilities of LLMs with the optimization abilities of EAs to optimize prompts\cite{b9}. However, the degree of optimization is limited because only a single evolutionary operator is used. FunSearch leverages the code generation capabilities of LLMs and the evaluation function to solve the ``hallucination'' problem of LLMs as well as generate effective heuristic algorithms, which provides a good example for solving combinatorial optimization problems using heuristic algorithms\cite{b10}. Nevertheless, FunSearch requires significant computational resource support, and millions of tokens are needed for a single task. EoH combines the semantic capturing and code generation capabilities of LLMs, using five types of evolutionary operators to guide LLMs to generate heuristic ideas and corresponding algorithms spontaneously\cite{b3}.

\section{Method}
\label{3}
\noindent
This section introduces the detailed implementation of the task scheduling scheme based on EC and heuristic algorithms.

\subsection{Overall framework}
\label{3.1}
\noindent
The edge server task scheduling algorithm based on EoH consists of heuristic strategies, a generator, and an evaluator, as shown in Fig.\ref{fig1}. 

Our heuristic strategies mimic genetic mutation and recombination processes in biological evolution, which can be divided into initialization, mutation, and evolution strategies. The initialization strategy generates the initial population. Mutation strategies generate offspring that are completely different from their parents. Evolution strategies generate better offspring than their parents while maintaining their basic characteristics. 

The generator consists of two phases. First, it generates heuristic descriptions based on heuristic strategies, leveraging LLMs' semantic capturing and natural language generation capabilities. Then, using LLMs' code generation capabilities, the corresponding code is spontaneously generated based on the heuristic descriptions, as shown in Fig. \ref{fig6}. The generated code is used for task selection.

\begin{figure}[t]
  \centering
  \includegraphics[width=0.47\textwidth]{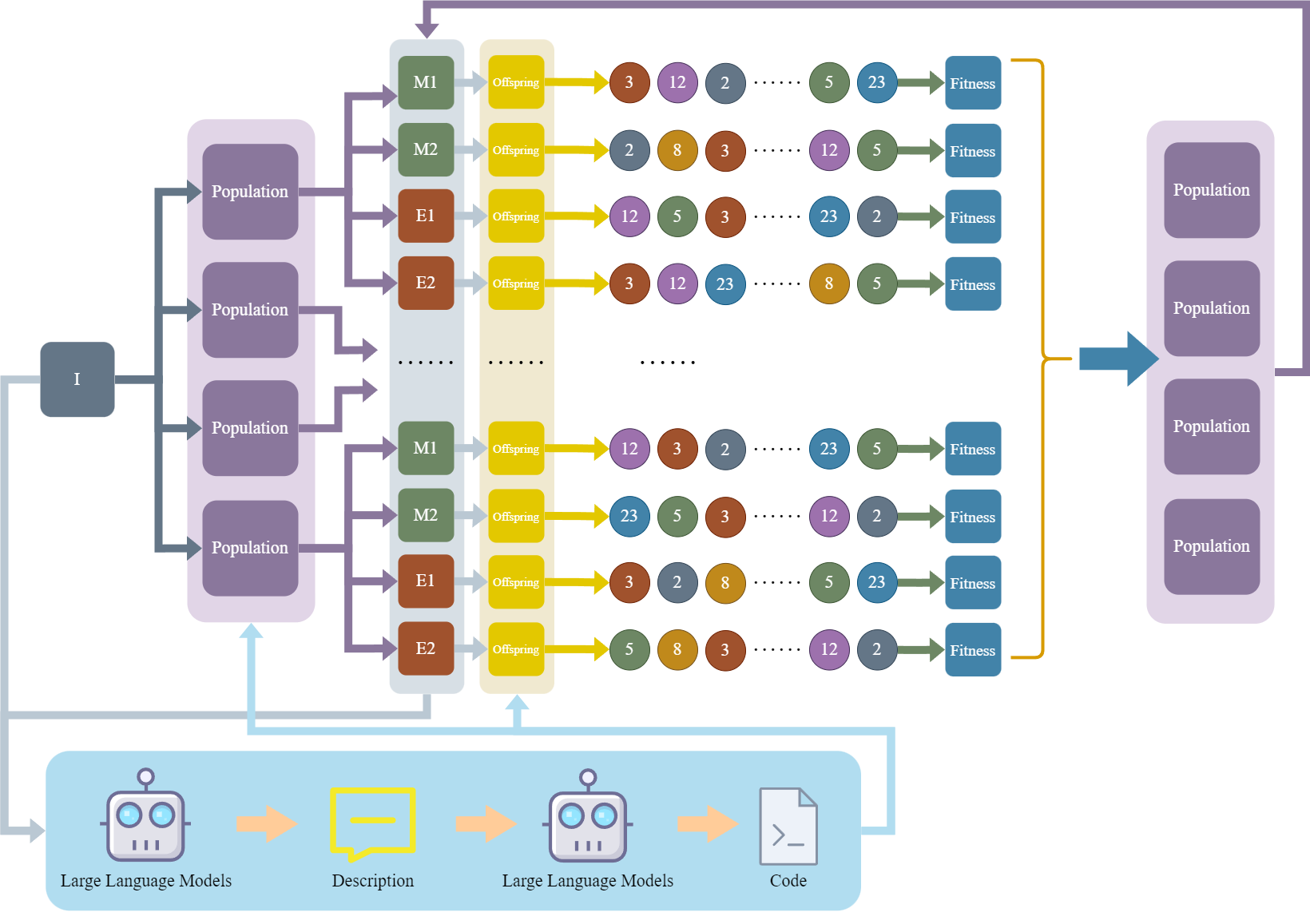} 
  \caption{The process of algorithm evolution.}
  \label{fig2}
\end{figure}

The evaluator first scores the tasks based on the generated algorithms and selects the task with the highest score for scheduling, iterating until all tasks are scheduled. It then evaluates the scheduling sequence to determine the fitness of the scoring algorithm for the task sequence. The fitness level decides whether the algorithm can proceed to the next round of evolution. The evolutionary process is illustrated in Fig. \ref{fig2}. 

\subsection{The design of heuristic strategies}
\label{3.2}
\noindent
This paper designs three heuristic strategies: initialization, mutation, and evolution. To ensure the uniqueness of the heuristic descriptions, the system performs a redundancy check every time a new heuristic strategy is applied to generate a description, except for the initialization strategy. If the generated heuristic description is found to be too similar to the existing ones, the system will execute the strategy again until a unique heuristic description is produced.

\subsubsection{The initialization strategy}
\label{3.2.1}
\noindent
The role of the initialization strategy is to create the initial population, which is the first step of running the heuristic algorithm successfully. This strategy guides the large-scale pre-trained LLMs in generating the initial heuristic descriptions using prompt engineering. These descriptions are the starting point for the algorithm's processing and form the basis for subsequent mutation and evolution strategies. The prompt templates are designed to provide a clear and effective framework to steer the model in generating the initial heuristics in the desired direction. Fig. \ref{fig6} shows two examples of prompt engineering.



\subsubsection{The mutation strategies}
\label{3.2.2}
\noindent
The mutation strategies in our heuristic algorithm are employed to inspire innovation by introducing new changes to promote diversity within the solution set. The key is to explore and generate new heuristic descriptions that are as different as possible from the original description, thus driving the evolution of heuristics.

\textbf{Mutation strategy M1: direct mutation.}

Mutation strategy M1 leads to direct generation, aiming to create offspring heuristics that exhibit obvious differences from their parent heuristics. This approach ensures that each mutation introduces new elements, driving the algorithm to explore new solutions.

\textbf{Mutation strategy M2: mutation remaining core idea.}

Unlike the M1 strategy, the mutation strategy M2 first identifies and extracts the core idea of the parent heuristic. Based on this core idea, it generates offspring heuristics that are the same in core concept but different in expression. This approach enhances the solution by adding diversity to the expression while keeping its intrinsic value.


\subsubsection{The evolution strategies}
\label{3.2.3}
\noindent
Evolution strategies are used to refine and optimize existing heuristics to enhance performance continuously. The design of these strategies ensures that the algorithm retains the advantages of the original heuristic while exploring new heuristics.

\textbf{Evolution strategy E1: performance optimization.}

The core idea of the evolutionary strategy E1 is to generate offspring heuristics that perform better than the parent heuristic. This strategy focuses on enhancing the solution's overall performance by introducing new algorithmic components or adjusting the implementation of existing logic so that the offspring heuristics can achieve higher efficiency and accuracy when handling the same tasks.


\textbf{Evolution strategy E2: parameters tuning.}

Evolutionary strategy E2 generates more effective offspring heuristics by tuning the parameters of the parent heuristic. This strategy focuses on adjustments for details, such as modifying parameters or reallocating their weights, to further refine and improve the heuristic description.


\subsection{The generator and evaluator}
\label{3.3}

\subsubsection{Code Generation}
\label{3.3.1}
\noindent
The main task of this section is to use the heuristic strategies defined in Section \ref{3.2} to generate executable Python code. Firstly, heuristic descriptions are generated according to the heuristic strategy. Then, these descriptions will be used as input, turning into heuristic algorithms through prompt rewriting. The specific steps in this phase are as follows:
\begin{itemize}
\item Step 1 Heuristic description generation: Based on the heuristic strategies in Section \ref{3.2}, generate heuristic descriptions that concisely and efficiently represent the details of the expected algorithm.
\item Step 2: Implement a novel prompt rewriting methodology to comprehensively analyze and refine the heuristic descriptions. This process ensures alignment with the specific requirements of the algorithm's implementation.
\item Step 3 Code Generation: Implement the algorithm function based on the rewritten prompt.
\end{itemize}


\subsubsection{Evolutionary Evaluation}
\label{3.3.2}
\noindent
Firstly, process the service request queue of the edge server, retaining the required data, which includes CPU, I/O, bandwidth, memory resources, the arrival time, execution time, and optional servers:
\begin{equation}
   q_i = (c_i, o_i, b_i, m_i, T_i, t_i, S_i)
\end{equation}

Model all service requests as a sequence$RS$:
\begin{equation}
   RS = \{q_1, q_2, ..., q_n\}
\end{equation}

Edge servers can be represented by set $S$:
\begin{equation}
    S = \{ s_1, s_2, ..., s_m\}
\end{equation}

Available resources for edge servers include CPU, I/0, bandwidth, and memory.
\begin{equation}
    s_j = (c_j, o_j, b_j, m_j)
\end{equation}

The task scheduling problem of the edge server can be simplified by reorganizing the request sequence \(RS\). Specifically, it involves determining the optimal reorganization of the request sequence \(RS\) given the available resources \(s_j\) to achieve optimal scheduling. This paper addresses two key scheduling optimization objectives: maximizing resource utilization and minimizing running time.

The task scheduling decision can be made according to the executable code generated in Section \ref{3.3.1}. The generated scoring function scores each task whose resource demand is satisfied. Then, the scoring sequence $S(RS)$ will be obtained.
\begin{equation}
    S(RS) = \{s(q_1), s(q_2), ..., s(q_N)\}
\end{equation}

Select the largest element $S(q_i)$in set $S(RS)$:
\begin{equation}
    s(q_i) = max\{S(RS)\}
\end{equation}

Add it to the schedule sequence, iterating until the request sequence is empty. Finally, the schedule sequence $SS$ will be obtained:
\begin{equation}
    SS = \{q_a, q_b, ..., q_n\}
\end{equation}

Subscript $a, b,... n$ is the subscript of the request in the request sequence$RS$. Then, evaluate the scheduling sequence $SS$ regarding resource utilization and running time. Generally, our algorithm will perform best when the evaluation coefficients $\alpha, \beta$ are set to 150 and 1.
\begin{equation}
    fitness = \alpha * avg(u) - \beta * avg(r),
\end{equation}
where $avg(u)$ is the average resource utilization rate, $avg(r)$ is the average running time, the calculation formula is as follows:
\begin{eqnarray}
    avg(u) = \frac{1}{m}\sum_{j = 1}^{m} max(c_j, o_j, b_j, m_j)\\avg(r) = \frac{1}{m}(finish\_time - start\_time)
\end{eqnarray}

\subsection{Algorithm Evolution}
\label{3.4}

\noindent
We adopted an EC-based strategy to optimize the task-scheduling sequence when designing efficient algorithms. Initially, we constructed a population composed of $N$ individuals, each representing a potential task scheduling scheme, and subsequently generated $4N$ offspring in each round of evolution. This process mimicked natural selection and genetic mutation mechanisms, allowing superior scheduling schemes to be preserved and evolved. 

In each round of evolution, the scoring scheme represented by each offspring will assign scores to the allocated tasks, reflecting whether the edge server should process the task at that time. We will select the highest-scoring tasks for execution until all tasks have been scheduled. Then, the system will comprehensively evaluate the scheduling sequence to determine the fitness of each scheme. The fitness represents the degree of match between the scoring scheme and the task set; the higher the fitness, the more suitable the scheme is for the given task set.

At the end of each evolution, we select $N$ offspring with the highest suitability from the $4N$ produced offspring to enter into the next round of evolution. This selection process simulates the principle of survival of the fittest, ensuring that the algorithm can continuously self-optimize and gradually find more efficient task-scheduling schemes.

After several rounds of evolutionary iteration, we select the scheduling scheme with the highest fitness from all offspring. With the optimal task scheduling sequence, this scheme is considered the best scheduling result for a specific set of tasks. This evolution-based task scheduling algorithm can adapt flexibly to various tasks and environmental conditions and automatically discover and utilize the potential associations between tasks, thereby significantly enhancing scheduling efficiency and system performance.

\section{Experiment}
\label{4}
\noindent
This section validates the effectiveness of the edge server task scheduling method based on EoH through a series of experiments on real datasets. Firstly, we compare the scheduling results of four different LLMs services. Then, we select the best-performing LLMs and compare the performance of the TS-EoH algorithm with other task scheduling methods on three real datasets. Finally, we assess the effectiveness of four evolutionary schemes through ablation experiments.

\subsection{Datasets}
\label{4.1}
\noindent
The experiment selected three public datasets: Google Cluster Trace, Alibaba Cluster Trace, and EUA-dataset.

\textbf{Google Cluster Trace}\cite{b19} collects data from Google's edge-cloud collaboration system, documenting detailed information about task submission, scheduling decisions, system configuration, and resource utilization.

\textbf{Alibaba Cluster Trace}\cite{b20} originates from multiple edge servers in Alibaba's production cluster environment, including task and job information, resource requests and usage, scheduling events, and system configuration and runtime information of applications.

\textbf{EUA-dataset}\cite{b21} contains the geographical locations of 125 base stations in Melbourne CBD and the 816 mobile users around. Our experiments consider base stations as edge servers and mobile users as service requests.


\begin{figure*}[t]
  \centering
  \includegraphics[width=\textwidth]{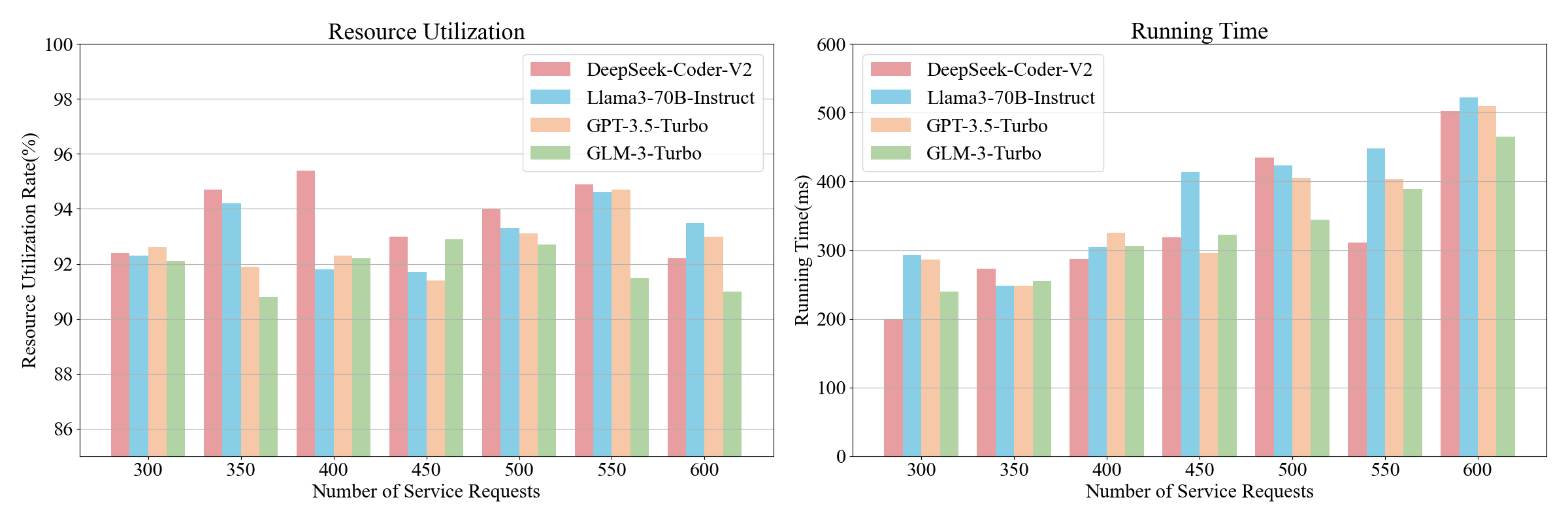} 
  \caption{Visual comparisons of different LLMs services.}
  \label{fig3}
  
\end{figure*}
\subsection{Comparison of LLMs}
\label{4.2}
\noindent
\textbf{Experimental settings}: To explore deeply and assess the scheduling performance of different LLMs, this experiment chooses four LLMs for comparison: GPT-3.5-Turbo, DeepSeek-Coder-V2, GLM-3-Turbo, and Llama3-70B-Instruct. The performance of these models in task scheduling on edge servers is compared within the Alibaba Cluster Trace dataset, across an interval of 300-600 task numbers, in terms of resource utilization rate and running time. These metrics reflect the different LLMs' capabilities in handling large-scale task scheduling issues and are key to evaluating algorithm performance. This experiment will not discuss the algorithm's efficiency as it is limited by LLMs' generation speed, which is related to uncontrollable external factors such as GPS access and inference side acceleration.

\textbf{Experimental results}: As shown in Fig. \ref{fig3}, different LLMs show significant variances in our task scheduling scheme related to their internal architecture and pre-training datasets. Regarding resource utilization, DeepSeek-Coder-V2 demonstrated an absolute advantage in more than 85\% of the task number range for its outstanding code-generating ability, followed by Llama3-70B-Instruct. As for running time, DeepSeek-Coder-V2 significantly reduced the overall runtime in most cases. Due to its concurrent processing capability, GLM-3-Turbo had a significantly shorter running time than the other three models when handling 600 tasks.

\subsection{Comparison of performance}
\label{4.3}
\noindent
\textbf{Experimental settings}: To comprehensively evaluate the superiority of TS-EoH, experiments are carried out on three datasets: Google Cluster Trace, Alibaba Cluster Trace, and EUA dataset. Experiments in Section \ref{4.2} demonstrate that DeepSeek-Coder-V2 best reflects the advantages of our algorithm; therefore, DeepSeek-Coder-V2 is selected as the base model for this section's experiments. In addition to our algorithm, we chose six current mainstream scheduling algorithms for comparative analysis: FCFS (First-Come, First-Served), HRRN (High-Response-Ratio-Next), Random, Greedy, ACO (Ant Colony Optimization), and RLPNet (Reinforcement Learning with Pointer Networks). The task number range is also 300-600.

\begin{itemize}
    \item FCFS\cite{b22}: Tasks are scheduled according to their arrival, with the first arriving tasks being executed.
    \item HRRN\cite{b23}: HRRN considers both the waiting time and service time of tasks, selecting the task with the highest response ratio for scheduling.
    \item Random\cite{b24}: Tasks are scheduled randomly by selecting the next task to execute without regard to any specific attributes or order.
    \item Greedy\cite{b25}: This method selects the currently optimal task to execute based on a certain optimization criterion.
    \item ACO\cite{b11}: A heuristic algorithm known for its natural parallelism and strong global search capability, effectively handling complex scheduling problems.
    \item RLPNet\cite{b4}: A service request scheduling algorithm based on deep reinforcement learning with pointer networks of multi-objective optimization.
\end{itemize}

\begin{figure*}[ht]
  \centering
  
  \begin{subfigure}[b]{\textwidth}
    \centering
    \begin{minipage}{\textwidth}
	\vspace{3pt}
	\centerline{\includegraphics[width=\textwidth]{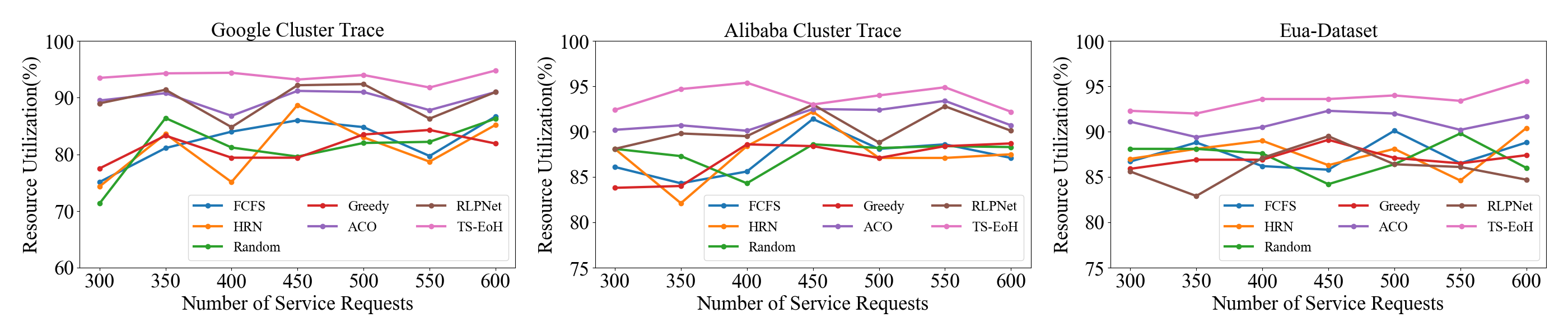}}
    \end{minipage}
    \caption{Resource Utilization}
    \label{fig:r}
  \end{subfigure}

  \begin{subfigure}[b]{\textwidth}
    \centering
    \begin{minipage}{\textwidth}
	\vspace{3pt}
	\centerline{\includegraphics[width=\textwidth]{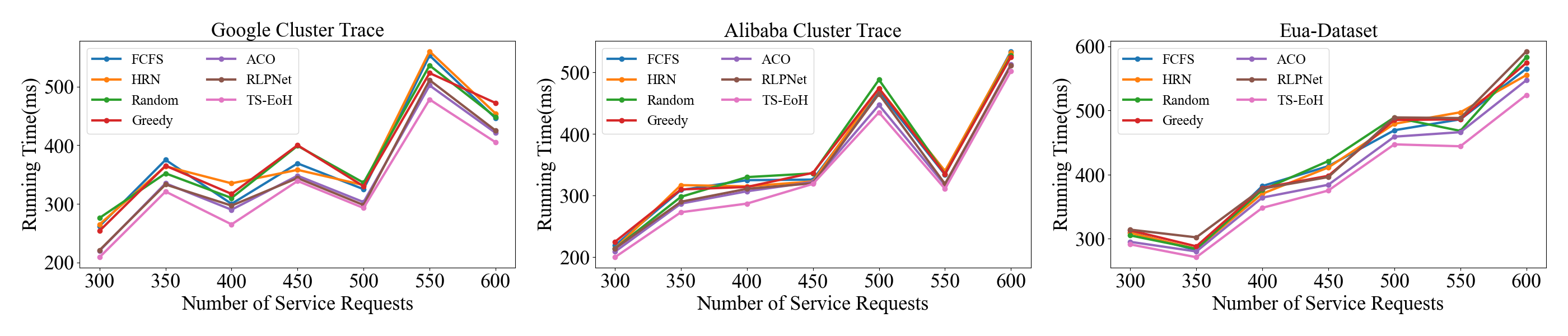}} 
    \end{minipage}
    \caption{Running Time}
    \label{fig:w}
  \end{subfigure}

  \caption{Visual comparison of performance on three datasets.}
  \label{fig4}
\end{figure*}

  

  
  

\textbf{Experimental results}: Our TS-EoH algorithm performs better than the other six task scheduling algorithms regarding resource utilization and runtime(Fig. \ref{fig4}). Regarding resource utilization rate, our TS-EoH algorithm significantly outperforms other algorithms. On the Google Cluster Trace, the maximum resource utilization reaches 94.8\%; on the Alibaba Cluster Trace, it is 95.4\%; and on the EUA dataset, it hits 95.6\%. Meanwhile, it is worth noticing that there's no obvious relationship between resource utilization rate and the number of tasks. Besides, As the number of tasks increases, the overall runtime tends to increase. However, due to the randomness in task selection, this trend is less apparent when the number of tasks does not differ significantly. Our EoH-based task scheduling algorithm can reduce task runtime to some extent, and the more tasks there are, the more apparent our algorithm's superiority is in running time. TS-EoH's resource utilization and running time superiority mainly lies in its self-evolution ability. At the same time, ACO may sacrifice its overall performance for local accuracy, and RLPNet has the problem of large state space, let alone the traditional task scheduling algorithms as FCFS, HRRN, etc.

\subsection{Ablation study}
\label{4.4}
\noindent
\textbf{Experimental settings}: A series of ablation experiments assessing the effects of the four heuristic strategies (M1, M2, E1, E2) are proposed in this study. These experiments aim to clarify the individual contributions of each strategy to the overall algorithm performance and explore potential synergistic effects among them by combining different evolutionary strategies. The experiments are categorized into three groups: single-strategy group (M1, M2, E1, E2), dual-strategy group (M1+M2, M2+E1, E1+E2, M1+E1, M2+E2, M1+E2), and triple-strategy group (M2+E1+E2, M1+E1+E2, M1+M2+E1). Each group of experiments utilizes DeepSeek-Coder-V2 and is conducted on the Alibaba Cluster Trace dataset with 500 tasks.

\begin{table}[h]
\caption{Ablation results of different strategy groups}
\label{tab7}
\begin{tabular}{p{0.16\textwidth} p{0.07\textwidth} p{0.10\textwidth} p{0.06\textwidth}}
\toprule[1.5pt]
Group & Strategy & Resource Utilisation Rate & Running Time\\
\midrule
Single-strategy Group & M1 & \textbf{92.3\%} & \textbf{408} \\
        & M2 & 92.5\% & 440 \\
        & E1 & 91.6\% & 445 \\
        & E2 & 89.2\% & 469 \\ 
        \midrule[1pt]
Dual-strategy Group & M1+M2 & \textbf{93.3\%} & \textbf{405} \\
         & M2+E1 & 91.4\% & 422 \\
         & E1+E2 & 87.1\% & 410 \\
         & M1+E2 & 90.0\% & 424 \\
         & M2+E2 & 91.7\% & 461 \\
         & M1+E1 & 92.2\% & 474 \\ 
         \midrule[1pt]
Triple-strategy Group & M2+E1+E2 & 90.8\% & \textbf{409} \\
         & M1+E1+E2 & 90.3\% & 457 \\
         & M1+M2+E1 & 91.7\% & 421 \\
         & M1+M2+E2 & \textbf{93.3\%} & 448 \\
\bottomrule[1.5pt]
\end{tabular}
\end{table}

\textbf{Experimental results}: Experiments indicate that mutation strategy M1 maximizes resource utilization rate and minimizes the running time to the greatest extent when using only one heuristic strategy. When employing two heuristic strategies, the synergistic effect of mutation strategies M1 and M2 significantly enhances the resource utilization rate and reduces running time. Under the combination of three strategies, approaches including two combined mutation strategies outperform those including two evolution strategies regarding resource utilization rate. In contrast, the method utilizing mutation strategy M2 and two evolution strategies excels in running time optimization.

\section{Conclusion}
\label{5}
\noindent
This paper proposes an innovative edge server task scheduling scheme based on the EoH framework to optimize resource utilization and service quality. We utilize heuristic algorithms and EC theories, coupled with the generative capabilities of LLMs, to seek the optimal scoring algorithm for efficient management and task scheduling on edge servers. Furthermore, through experiments, we explore our algorithm's most suitable LLM services and compare its superiority over existing state-of-the-art algorithms. We also clarify the impact of different evolutionary strategies on scheduling results. Future work could focus on further optimizing the mutation strategy to maximize its advantages in the evolutionary process. Additionally, it is meaningful to select several optimal populations through extensive experiments and generate instructional data through the SELF-INSTRUCT framework\cite{b26}, which can be applied to fine-tune the LLMs to produce results more closely aligned with the application scenarios outlined in this paper.

\bibliographystyle{IEEEtran}
\bibliography{ref}

\end{document}